\documentclass[12pt,a4paper]{article}

\usepackage[]{babel}
\usepackage[utf8x]{inputenc}
\usepackage{float}
\usepackage{amsmath}
\usepackage{graphicx}
\usepackage[colorinlistoftodos]{todonotes}
\usepackage{url}
\usepackage{hyperref}
\usepackage{array}
\usepackage{tabularx}
\usepackage{setspace}
\usepackage{abstract}
\usepackage[T1]{fontenc}
\usepackage[top=2cm, bottom=2cm, left=2cm, right=2cm]{geometry}
\usepackage{subfig}
\usepackage[arrowdel]{physics}
\usepackage{xcolor}
\usepackage{empheq}
\usepackage{comment}
\usepackage{cite} 
\usepackage{amsmath,bm}
\usepackage{authblk}

\hypersetup{							
pdfauthor = {Premier Auteur,
			Deuxième Auteur,
			Troisième Auteur,
    		Quatrième Auteur},			
pdftitle = {Nom du Projet -
			Sujet du Projet},			
pdfsubject = {Mémoire de Projet},		
pdfkeywords = {Tag1, Tag2, Tag3, ...},	
pdfstartview={FitH}}					

\begin{document}

\newcommand{\HRule}{\rule{\linewidth}{0.5mm}}

\setlength{\baselineskip}{1.5\baselineskip}

\pagenumbering{arabic}
\setcounter{page}{0}


\renewcommand{\arraystretch}{1.5}
\setcounter{page}{1}
\begin{center}
{\huge \bfseries Effect of temperature anisotropy on the dynamics of geodesic acoustic modes\\[0.4cm]}
J.N. Sama$^1$, A. Biancalani$^{2,3}$, A. Bottino$^3$, I. Chavdarovski$^4$, D. Del Sarto$^1$, A. Ghizzo$^1$, T. Hayward-Schneider$^3$, P. Lauber$^3$, B. Rettino$^3$ and F. Vannini$^3$\\
$^1$Institut Jean Lamour UMR 7198, Université de Lorraine-CNRS, Nancy, France\\
$^2$Léonard de Vinci Pôle Universitaire, Research Center, Paris La Défense, France\\
$^3$Max Planck Institute for Plasma Physics, Garching, Germany\\
$^4$Korea Institute of Fusion Energy, Daejeon, South Korea\\
\end{center}
\section*{Abstract}
In this work, we revisit the linear gyro-kinetic theory of geodesic acoustic modes (GAMs) and derive a general dispersion relation for an arbitrary equilibrium distribution function of ions. A bi-Maxwellian distribution of ions is then used to study the effects of ion temperature anisotropy on GAM frequency and growth rate. We find that ion temperature anisotropy yields sensible modifications to both the GAM frequency and growth rate as both tend to increase with anisotropy and these results are strongly affected by the electron to ion temperature ratio. 
\section{Introduction}
Geodesic acoustic modes (GAMs) \textcolor{blue}{\cite{winsor}}, are oscillating axisymmetric perturbations that are unique to configurations with closed magnetic field lines with a geodesic curvature, like tokamaks. They are the oscillating counterparts of the zero frequency zonal flow (ZFZF)  \textcolor{blue}{\cite{A}} and are examples of zonal structures. Zonal structures are of great interest to magnetic fusion reactors due to their potential capabilities of generating non-linear equilibrium \textcolor{blue}{\cite{chen}} by regulating microscopic turbulence and its associated heat and particle transport. 

GAMs have been largely studied in literature both analytically \textcolor{blue}{\cite{Zh,hassam,smolyakov,zonca1,JB,X,N,D,Z}},\textcolor{blue}{\cite{Y,M,H,GD}} and numerically \textcolor{blue}{\cite{Bc,V,I}}. A key aspect in the linear gyro-kinetic theory of GAMs is the determination of mode frequency and damping rate. The GAM frequency is of the order of the ion sound frequency and its major damping mechanism is collisionless damping. Analytical expressions of GAM frequency and growth rate, can be found for example in refs.\textcolor{blue}{\cite{sw,Zh}}. These expressions were obtained assuming Maxwellian distributions of ions and electrons with no temperature anisotropy.

Tokamak plasmas are generally modeled in analytical theory assuming isotropic Maxwellian distributions of ions and electrons. However in reality, there can be several sources of anisotropy in tokamak plasmas. Anisotropy in tokamaks can be introduced by auxiliary heating such as neutral beam injection (NBI), which can generate  a strong parallel temperature anisotropy, whereas strong perpendicular temperature anisotropy can be observed when using ion cyclotron resonance heating (ICRH). Parallel and perpendicular  here are defined with respect to the equilibrium magnetic field. Generally, ion temperature anisotropy, both gyrotropic and non-gyrotropic can be generated due to the action of the traceless rate of shear, which anisotropically heats the in-plane components of the pressure tensor by tapping kinetic energy from shear flow, when the local gradient of the ion fluid velocity, say $\omega\sim||{\bm\nabla}{\bm u}_i||$, is not negligible with respect to the local ion cyclotron frequency $\Omega_i$ \textcolor{blue}{\cite{DD}}. This condition is likely to occur in developed turbulence, since it can be verified on the vorticity sheets delimiting vortex structures \textcolor{blue}{\cite{DD1}}. In particular, as long as the ratio  $\omega/\Omega_i$, remains small enough, the generated anisotropy is mostly gyrotropic and thus compatible with a gyrokinetic description \textcolor{blue}{\cite{DD2}} and it can be thus related to the first order finite-Larmor radius corrections to double-adiabatic closures \textcolor{blue}{\cite{kau,thom,mac,cerri}} . K. Sasaki \textcolor{blue}{\cite{K}} measure reasonably high ion temperature anisotropy in EXTRAP-T2 and large pressure anisotropies have also been reported in refs. \textcolor{blue}{\cite{wz,mj}}.  H. Ren in ref.\textcolor{blue}{\cite{HR}} studied the impact of ion temperature anisotropy on GAM frequency and growth rate in the limit of a vanishing electron to ion temperature ratio, with ions described by a bi-Maxwellian distribution and found that ion temperature anisotropy modifies the linear dynamics of GAMs. However, the GAM dynamics is known to strongly depend on the electron to ion temperature ratio \textcolor{blue}{\cite{Zh,zonca1,Bc,sw}}. Hence a finite  electron to ion temperature ratio must be retained in a complete linear theory of GAMs. 

In this work, we investigate the linear dynamics of GAMs with a bi-Maxwellian distribution of ions and assuming adiabatic electrons as in Refs.\textcolor{blue}{\cite{zonca1,sw}}. We generalize the work in Ref. \textcolor{blue}{\cite{HR}}, to a general value of electron to ion temperature ratio and by keeping account of a gyro-tropic ion temperature anisotropy, using an approach based on the standard limit of small finite-orbit-radius and small finite-orbit-width, kept up to the leading order (consistently with Refs.\textcolor{blue}{\cite{zonca2,zonca1}}). We show that in the appropriate limits, we recover the GAM dispersion derived in Refs. \textcolor{blue}{\cite{HR,zonca1,JB}} from the general GAM dispersion relation which we here obtain. From our study, we find that the ion temperature anisotropy yields a sensible modification to both the real and imaginary part of the frequency, as both tend to be increasing functions of $\chi=\frac{T_{\perp,i}}{T_{||,i}}$, and this result is strongly affected by the electron to ion temperature ratio, $\tau=\frac{T_e}{T_i}$. The equivalent ion temperature $T_i$, is defined such that it corresponds to the same total pressure as that of the anisotropic distribution ($T_i=\frac{T_{||,i}}{3}+\frac{2T_{\perp,i}}{3}$). 

The first section of our work is an introduction and description of motivations for this work. In the second section, we derive a general linear GAM dispersion relation for an arbitrary distribution function. In section three, we solve the dispersion relation with a bi-Maxwellian distribution of ions and study the impact of ion temperature anisotropy and electron to ion temperature ratio on GAM frequency and growth rate. We apply our theory to an experimentally relevant case in section four and conclude in section five.

\section{The model for a general distribution function}

In this section, we use the gyro-kinetic formalism to study the physics of GAMs in the electrostatic limit. The fundamental equations of this model are the gyro-kinetic Vlasov (\textcolor{blue}{\ref{vlasov}}) and Poisson equations (\textcolor{blue}{\ref{quasi}}).
\begin{equation}
    \frac{\partial f_s}{\partial t}+\dot{\vec{R}}\cdot\frac{\partial f_s}{\partial\vec{R} }+\dot{E}\frac{\partial f_s}{\partial E }=0
    \label{vlasov}
\end{equation}
\begin{equation}
    -\grad\cdot\left(\frac{n_{0,i}m_ic^2}{B^2}\nabla_{\perp}\Phi\right)=\int dW_iZ_ieJ_{0,i}f_i-\int dW_eef_e
    \label{quasi}
\end{equation}
where $f_s$, is the distribution function of a given species, $\vec{R}$ is the particle position vector, $E=\frac{m_s}{2}\left(v_{||}^2+v_\perp^2\right)$, is the particle energy, $e$, the electron charge, $n_{0,s}$, the equilibrium density of a species $s$, $m$, particle mass, $c$, speed of light, $B$, magnitude of magnetic field, $Z_s$, species charge number, $J_{0,i}$, ion gyro-average operator and $dW_s$, is the volume element in velocity space.

We make the following assumptions,
\begin{enumerate}
    \item We consider adiabatic electrons.
    \item Neglect magnetic fluctuations.
    \item Use flat density and temperature profiles.
\end{enumerate}

\subsection{Linear analysis}

We linearize the Vlasov and quasi-neutrality equation by splitting each quantity into an equilibrium and perturbed component such that,
\begin{equation}
    f_s=f_{0,s}+f_{1,s}
    \label{f}
\end{equation}
\begin{equation}
    \dot{\vec{R}}=\dot{\vec{R}}_0+\dot{\vec{R}}_1
    \label{r}
\end{equation}
\begin{equation}
    \dot{E}=\dot{E}_0+\dot{E}_1
    \label{e}
\end{equation}
\begin{equation}
    \Phi=\Phi_1
    \label{p}
\end{equation}
Where $\dot{\vec{R}}_0$ is the unperturbed particle velocity $i.e$ $\dot{\vec{R}}_0= \vec{v}_{||}+\vec{v}_{\nabla B}+\vec{v}_{curv B} $
and $E_0=\frac{m(v_{||}^2+v_{\perp}^2)}{2}$.
\subsection*{Linear Vlasov equation}
Substituting equation (\textcolor{blue}{\ref{f}})-(\textcolor{blue}{\ref{p}}) in (\textcolor{blue}{\ref{vlasov}}), and neglecting second and higher order terms, the linear Vlasov equation reads,
\begin{equation}
    \frac{\partial f_{1,s}}{\partial t}+\dot{\vec{R}}_0\cdot\frac{\partial f_{1,s}}{\partial\vec{R} }=-\dot{E}_1\frac{\partial f_{0,s}}{\partial E }
    \label{lvlasov}
\end{equation}
This linear Vlasov equation can be further simplified by splitting the perturbed distribution function into an adiabatic and non-adiabatic component,
\begin{equation}
    f_{1,s}=E_1\frac{\partial f_{0,s}}{\partial E}+h_s
    \label{adia}
\end{equation}
Substituting equation (\textcolor{blue}{\ref{adia}}) in (\textcolor{blue}{\ref{vlasov}}), we obtain the equation for the non-adiabatic part of the perturbed distribution function,
\begin{equation}
    \frac{\partial h_s}{\partial t}+\dot{\vec{R}}_0\cdot\frac{\partial h_s}{\partial\vec{R} }=-\dot{E}_1\frac{\partial f_{0,s}}{\partial E }
    \label{avlasov}
\end{equation}
Using the expression of the equilibrium velocity and perturbations of the form $x \longrightarrow \exp{ik_r-i\omega t}+c.c$, in equation (\textcolor{blue}{\ref{avlasov}}) we obtain the expression,
\begin{equation}
\left(\omega_{t,s}\frac{\partial}{\partial \theta}-i(\omega+\omega_{d,s})\right)h_s=i\omega E_1\frac{\partial f_{0,s}}{\partial E}
\end{equation}
        Where $\omega_t=\frac{v_{||}}{qR_0}$, $q$ is the safety factor, $\omega_{d,s}=\bar{\omega}_{d,s}\sin{\theta}$ and $\bar{\omega}_{d,s}=\frac{cm_sk_r}{Z_seB_0R_0}\left(v_{||}^2+\frac{v_{\perp}^2}{2}\right)$, $E_1=Z_seJ_{0,s}\Phi_1$ and $k_r$ is the radial wave number . Since GAMs are predominantly zonal, we can further divide the non-adiabatic part of the perturbed distribution function into a zonal and non-zonal part,
\begin{equation}
    h=\Bar{h}+\delta h
\end{equation}
similarly, we write the scalar potential as,
\begin{equation}
    \Phi_1=\Bar{\Phi}+\Tilde{\Phi} 
\end{equation}
where the overhead bar represent the zonal components. Using these definitions, and making a flux surface average of the gyro-kinetic equation to eliminate the zonal component of the non-adiabatic part of the perturbed distribution function, the linear Vlasov equation reduces to,
\begin{equation}
\left(\omega_{t,s}\frac{\partial}{\partial \theta}-i(\omega+\omega_{d,s})\right)\delta h_s=iZ_s\frac{\partial f_{0,s}}{\partial E}\left(\omega J_{0,s}\Tilde{\Phi}-\omega_{d,s} J_{0,s}\Bar{\Phi}\right)
\label{tvlasov}
\end{equation}
The corresponding vorticity equation is obtained by multiplying this relation by the gyro-average operator and integrating over the velocity space.
\begin{equation}
\left<J_{0,s}\left(\omega_{t,s}\frac{\partial}{\partial \theta}-i(\omega+\omega_{d,s})\right)\delta h_s\right>_W=\left<iZ_s\frac{\partial f_{0,s}}{\partial E}\left(\omega J_{0,s}^2\Tilde{\Phi}-\omega_{d,s} J_{0,s}^2\Bar{\Phi}\right)\right>_W
\label{vorti}
\end{equation}
Considering all the changes of variable we have made, the perturbed distribution function has the form,
\begin{equation}
    f_{1,s}=Z_seJ_{0,s}\frac{\partial f_{0,s}}{\partial E}\Tilde{\Phi}+\delta h_s
\end{equation}
\subsection*{Linear quasi-neutrality equation}
Using a similar approach, we substitute equations (\textcolor{blue}{\ref{f}}) and (\textcolor{blue}{\ref{p}}) into (\textcolor{blue}{\ref{quasi}}) and simplify to obtain the equation below,
\begin{equation}
    \frac{m_ic^2}{B^2}k_r^2\bar{\Phi}=e^2\left<J_{0,i}^2\frac{\partial f_{0,i}}{\partial E}\right>_W\left(1+\frac{\left<J_{0,e}^2\frac{\partial f_{0,e}}{\partial E}\right>_W}{\left<J_{0,i}^2\frac{\partial f_{0,i}}{\partial E}\right>_W}\right)\Tilde{\Phi}+e\left<J_{0,i}\delta h_i\right>_W
\end{equation}
where $\left<..\right>_W$, represent the integral over velocity space. The non-adiabatic part of the perturbed electron distribution function has been neglected in accordance with our assumptions. 
\subsection{Ordering of the gyro-kinetic equation}
The gyro-kinetic equation (\textcolor{blue}{\ref{tvlasov}}) describes a wide range of phenomena at different time scales. In order to study GAMs, we need to apply an appropriate ordering that will filter out irrelevant time scales in the dynamics. The GAM frequency is of the order of ion sound frequency. The ordering is done by comparing this frequency with the characteristic frequencies in our system $i.e$ $ \omega_{t,s}$, $\omega_{d,s}$.
\begin{align}
    \frac{\omega_{t,i}}{\omega}& \sim o(1)& \frac{\omega_{d,i}}{\omega}& \sim o(\epsilon)&\frac{\delta h_i}{f_{0,i}}& \sim o(\epsilon)&
\end{align}
\begin{align}
    \frac{\omega_{t,e}}{\omega}& >> 1 & \frac{\omega_{d,e}}{\omega}& \sim o(1)&\frac{\delta h_e}{f_{0,e}}& \sim o(\epsilon)&
\end{align}
In the leading order, the ion and electron gyro-kinetic equations are respectively,
\begin{equation}
    \left(\frac{\omega_{t,i}}{\omega}\frac{\partial}{\partial \theta}-i\right)\delta h_i=iZ_ie\frac{\partial f_{0,i}}{\partial E}\left(J_{0,i}\Tilde{\Phi}-J_{0,i}\frac{\omega_{d,i}}{\omega}\Bar{\Phi}\right)
    \label{rvlasov}
\end{equation}

\begin{equation}
\delta h_e=0
\end{equation}
\subsection{General form of dispersion relation}
We consider the following form for the non-zonal perturbed ion distribution function and scalar potential. 
\begin{equation}
    \Tilde{\Phi}=\Tilde{\Phi}_s\sin{\theta}+\Tilde{\Phi}_c\cos{\theta}
\end{equation}
\begin{equation}
    \delta h_i=\delta h_{i,s}\sin{\theta}+\delta h_{i,c}\cos{\theta}
\end{equation}
Substituting these relations into equation (\textcolor{blue}{\ref{rvlasov}}) and separating the sine and cosine components, we obtain,
\begin{equation}
    \delta h_{i,s}=\frac{iZ_iJ_{0,i}\frac{\partial f_{0,i}}{\partial E}}{\left(\frac{\omega_{t,i}}{\omega}\right)^2-1}\left[-i\left(\Tilde{\Phi}_s-\left(\frac{\Bar{\omega}_{d,i}}{\omega}\right)\Bar{\Phi}\right)+\frac{\omega_{t,i}}{\omega}\Tilde{\Phi}_c\right]
    \label{ks}
\end{equation}
\begin{equation}
    \delta h_{i,c}=-\frac{iZ_iJ_{0,i}\frac{\partial f_{0,i}}{\partial E}}{\left(\frac{\omega_{t,i}}{\omega}\right)^2-1}\left[\frac{\omega_{t,i}}{\omega}\left(\Tilde{\Phi}_s-\left(\frac{\Bar{\omega}_{d,i}}{\omega}\right)\Bar{\Phi}\right)+i\Tilde{\Phi}_c\right]
    \label{kc}
\end{equation}
Following the same procedure with the quasi-neutrality equation,
\begin{equation}
    \Tilde{\Phi}_c=-\frac{\left<J_{0,i}\delta h_{i,c}\right>_W}{e\left<J_{0,i}^2\frac{\partial f_{0,i}}{\partial E}\right>_W\left(1+\frac{\left<J_{0,e}^2\frac{\partial f_{0,e}}{\partial E}\right>_W}{\left<J_{0,i}^2\frac{\partial f_{0,i}}{\partial E}\right>_W}\right)}
    \label{pc}
\end{equation}
\begin{equation}
    \Tilde{\Phi}_s=-\frac{\left<J_{0,i}\delta h_{i,s}\right>_W}{e\left<J_{0,i}^2\frac{\partial f_{0,i}}{\partial E}\right>_W\left(1+\frac{\left<J_{0,e}^2\frac{\partial f_{0,e}}{\partial E}\right>_W}{\left<J_{0,i}^2\frac{\partial f_{0,i}}{\partial E}\right>_W}\right)}
    \label{ps}
\end{equation}
Taking the flux surface average of the quasi-neutrality equation (\textcolor{blue}{\ref{rvlasov}}) and the vorticity equation (\textcolor{blue}{\ref{vorti}}) we have respectively,
\begin{equation}
    \frac{m_ic^2}{B^2}k_r^2\bar{\Phi}=e\overline{\left<J_{0,i}\delta h_i\right>}_W
    \label{11}
\end{equation}
\begin{equation}
\overline{\left<J_{0,i}\delta h_i\right>}_W=-\frac{\overline{\left<J_{0,i}\omega_{d,i}\delta h_i\right>}_W}{\omega}
\label{12}
\end{equation}
 Substituting (\textcolor{blue}{\ref{12}}) into (\textcolor{blue}{\ref{11}}) and evaluating the flux surface average we obtain,
\begin{equation}
    \frac{2m_i}{B^2}k_r^2\bar{\Phi}=-\frac{e}{c^2\omega}\left<J_{0,i}\bar{\omega}_{d,i}\delta h_{i,s}\right>
\end{equation}
Substituting equation (\textcolor{blue}{\ref{ks}}), we obtain the general dispersion relation of GAMs in the leading order,
\begin{equation}
    \frac{2m_ic^2k_r^2}{B^2}\bar{\Phi}=-e^2\omega\left<\frac{J_{0,i}^2\omega_{d,i}\frac{\partial f_{0,i}}{\partial E}}{\omega_{t,i}^2-\omega^2}\right>_W\Tilde{\Phi}_s+e^2\left<\frac{J_{0,i}^2\omega_{d,i}^2\frac{\partial f_{0,i}}{\partial E}}{\omega_{t,i}^2-\omega^2}\right>_W\Bar{\Phi}
    \label{disrela}
\end{equation}
\section{Case of bi-Maxwellian }
\subsection{Derivation of dispersion relation}
To evaluate the integrals over velocity space given in the general linear GAM dispersion relation (\textcolor{blue}{\ref{disrela}}), we have to choose an equilibrium distribution function for ions. In this work, we consider a bi-Maxwellian as our equilibrium distribution function for ions, while  a regular Maxwellian is used for electrons. It is normalised such that its integral over velocity space equals one ($n_{0,i}=1$). We take $J_{0,i}=1$ (drift kinetic limit).
\begin{equation}
    f_{0,i}=\left(\frac{m_i}{2\pi T_{||}}\right)^\frac{1}{2}\left(\frac{m_i}{2\pi T_\perp}\right)\exp{-\frac{m_i}{2}\left(\frac{v_{||}^2}{T_{||}}+\frac{v_\perp^2}{T_\perp}\right)}
\end{equation}
By defining an equivalent temperature, $T_i=\frac{T_{||,i}}{3}+\frac{2T_{\perp,i}}{3}$, the bi-Maxwellian can be written in the form,
\begin{equation}
    f_{0,i}=\frac{ b^\frac{3}{2}}{\pi^\frac{3}{2}v_t^3\chi}\exp{-b\left(\frac{v_{||}^2+v_\perp^2\chi^{-1}}{v_t^2}\right)}
\end{equation}
Where $b=\frac{2\chi+1}{3}$ and $\chi=\frac{T_{\perp,i}}{T_{||,i}}$, $v_t=\sqrt{\frac{2T_i}{m}}$ and $E=\frac{m}{2}\left(v_{||}^2+v_\perp^2\right)$.

we have,
\begin{equation}
    \frac{\partial f_{0,i}}{\partial E}=-\frac{b}{T_i}f_{0,i}
\end{equation}
Equation (\textcolor{blue}{\ref{ps})} then reduces to,
\begin{equation}
  \Tilde{\Phi}_s=\frac{\omega\left<\frac{\bar{\omega}_{d,i}f_{0,i}}{\omega_t^2-\omega^2}\right>_W}{1+\frac{1}{\tau b}+\omega^2\left<\frac{f_{0,i}}{\omega_t^2-\omega^2}\right>_W}\bar{\Phi}  
\end{equation}
Where $\tau=\frac{T_e}{T_i}$. Substituting this result in the general dispersion relation we have,
\begin{equation}
  \frac{2m_ic^2k_r^2}{B^2}+\frac{e^2b}{T_i}\left[\left<\frac{\bar{\omega}_{d,i}^2f_{0,i}}{\omega_t^2-\omega^2}\right>_W-\frac{\omega^2\left<\frac{\bar{\omega}_{d,i}f_{0,i}}{\omega_t^2-\omega^2}\right>_W^2}{1+\frac{1}{\tau b}+\omega^2\left<\frac{f_{0,i}}{\omega_t^2-\omega^2}\right>_W}\right]=0  
  \label{13}
\end{equation}
 These three integrals evaluate to,
\begin{equation}
    \left<\frac{\bar{\omega}_{d,i}f_{0,i}}{\omega_t^2-\omega^2}\right>_W=\frac{cm_ik_rv_t^2}{eB_0R_0b\omega_0^2y}\left[y+\left(\frac{\chi}{2}+y^2\right)Z(y)\right]
\end{equation}
\begin{equation}
  \left<\frac{\bar{\omega}_{d,i}^2f_{0,i}}{\omega_t^2-\omega^2}\right>_W=\left(\frac{cm_ik_rv_t^2}{eB_0R_0b}\right)^2\frac{1}{\omega_0^2y}\left[\frac{y}{2}+y^3+y\chi+\left(\frac{\chi^2}{2}+\chi y^2+y^4\right)Z(y)\right]
\end{equation}
\begin{equation}
   \left<\frac{f_{0,i}}{\omega_t^2-\omega^2}\right>_W= \frac{1}{\omega_0^2}\frac{Z(y)}{y}
\end{equation}
Where $Z(y)$ is the plasma dispersion function and,
\begin{align*}
    y&=\frac{\omega}{\omega_0}& \omega_0&=\frac{v_t}{qR_0\sqrt{b}}
\end{align*}
    Then equation (\textcolor{blue}{\ref{13}}) becomes,
\begin{equation}
    y+q^2\left[F(y)-\frac{N^2(y)}{D(y)}\right]=0
    \label{real}
\end{equation}
With,
\begin{equation}
    F(y)=\frac{y}{2}+y^3+y\chi+\left(\frac{\chi^2}{2}+y^2\chi+y^4\right)Z(y)
    \label{f1}
\end{equation}
\begin{equation}
    N(y)=y+\left(\frac{\chi}{2}+y^2\right)Z(y)
    \label{n1}
\end{equation}
\begin{equation}
    D(y)=\frac{1}{y}\left(1+\frac{1}{\tau b}\right)+Z(y)
    \label{d1}
\end{equation}
\subsection{Comparison with published results}
    In the limit $\chi=1$, we recover the GAM dispersion relations in \textcolor{blue}{\cite{zonca1,zonca2,JB}}. The dispersion relation (\textcolor{blue}{\ref{real}}) can be recasted in the form below by substituting equations (\textcolor{blue}{\ref{f1}}),(\textcolor{blue}{\ref{n1}}),(\textcolor{blue}{\ref{d1}}) in (\textcolor{blue}{\ref{real}}).
\begin{equation}
    \frac{1}{b\tau}\left[\frac{1}{q^2}+\frac{1}{2}+\chi+y^2+\left(y^3+y\chi+\frac{\chi^2}{2y}\right)Z(y)\right]+\left[\frac{1}{q^2}+\frac{1}{2}+\chi+\frac{yZ(y)}{q^2}+\left(\frac{y}{2}+y\chi+\frac{\chi^2}{2y}\right)+\frac{\chi^2}{4}{Z(y)}^2\right]=0
    \label{ddd}
    \end{equation}
\subsection*{Case $\tau$ $\rightarrow$ $0$}
In this limit, the first term in the square bracket is large compared to the second term. Neglecting this second term, we recover the dispersion relation in ref. \textcolor{blue}{\cite{HR}}.
\begin{equation}
    \frac{1}{q^2}+\frac{1}{2}+\chi+y^2+\left(y^3+y\chi+\frac{\chi^2}{2y}\right)Z(y)=0
\end{equation}
The plots below shows the frequency and growth rate obtained from the complete dispersion relation (\textcolor{blue}{\ref{real}}) and that obtained in the $\tau\rightarrow 0$ limit.  
\begin{figure}[H]
    \centering
    \subfloat[\centering]{{\includegraphics[width=8.4cm]{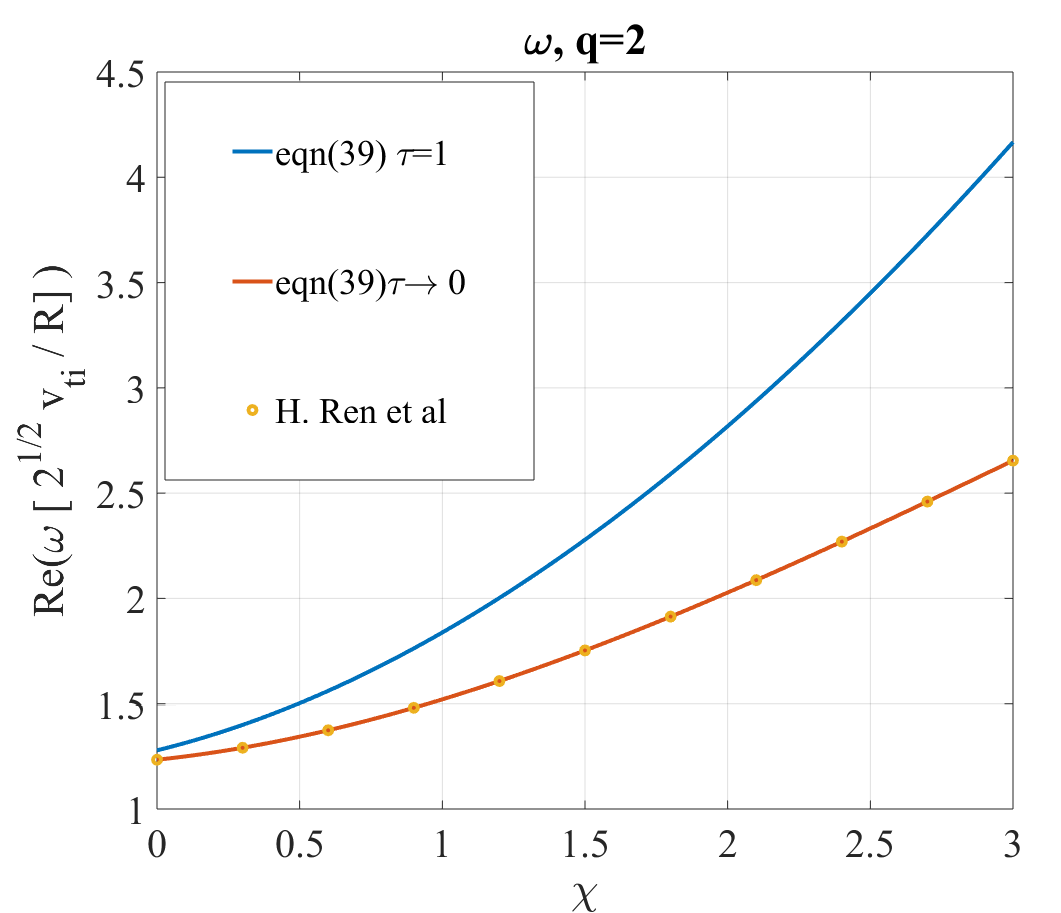}}}
    \subfloat[\centering]{{\includegraphics[width=9.0cm]{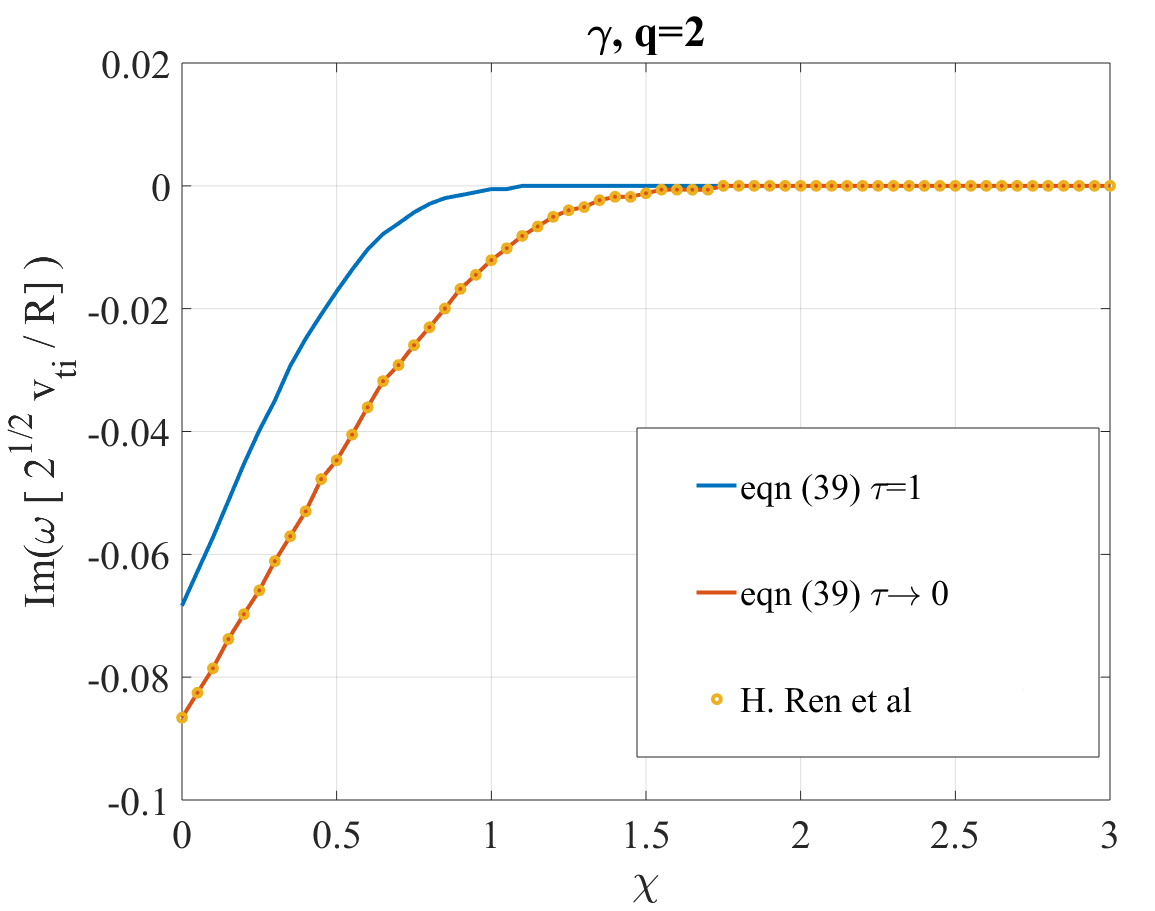}}}
    \caption{(a)Frequency (b)Growth rate}
\end{figure}
\subsection*{Case $\tau\rightarrow\infty$}
In this limit, the terms in the second square bracket in equation(\textcolor{blue}{\ref{ddd}}) are larger compare to those in the first. So the dispersion relation reduces to,
\begin{equation}
    \frac{1}{q^2}+\frac{1}{2}+\chi+\frac{yZ(y)}{q^2}+\left(\frac{y}{2}+y\chi+\frac{\chi^2}{2y}\right)+\frac{\chi^2}{4}{Z(y)}^2=0
\end{equation}
The plots below shows the frequency and growth rate obtained from the complete dispersion relation (\textcolor{blue}{\ref{real}}) and that obtained in the $\tau\rightarrow\infty$ limit.
\begin{figure}[H]
    \centering
    \subfloat[\centering]{{\includegraphics[width=9.5cm]{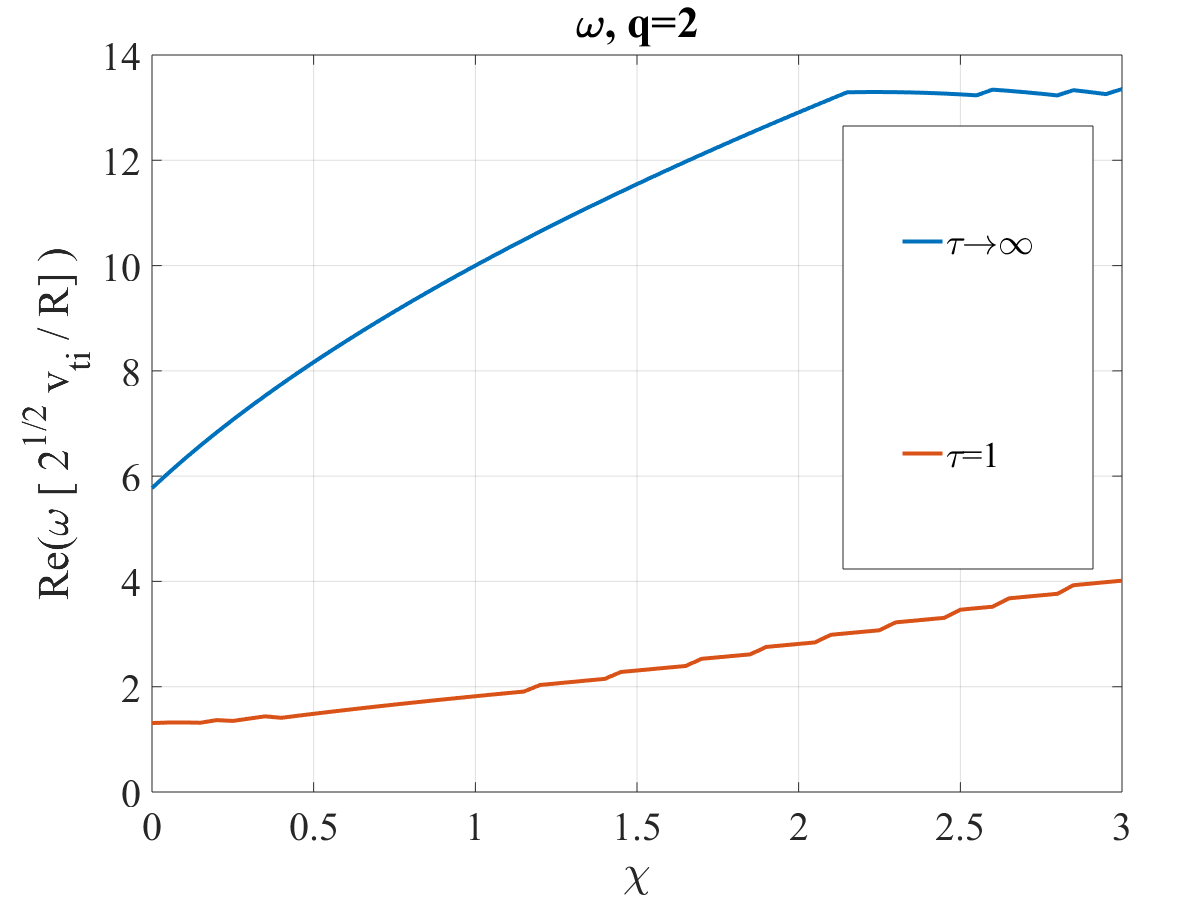}}}
    \subfloat[\centering]{{\includegraphics[width=9.5cm]{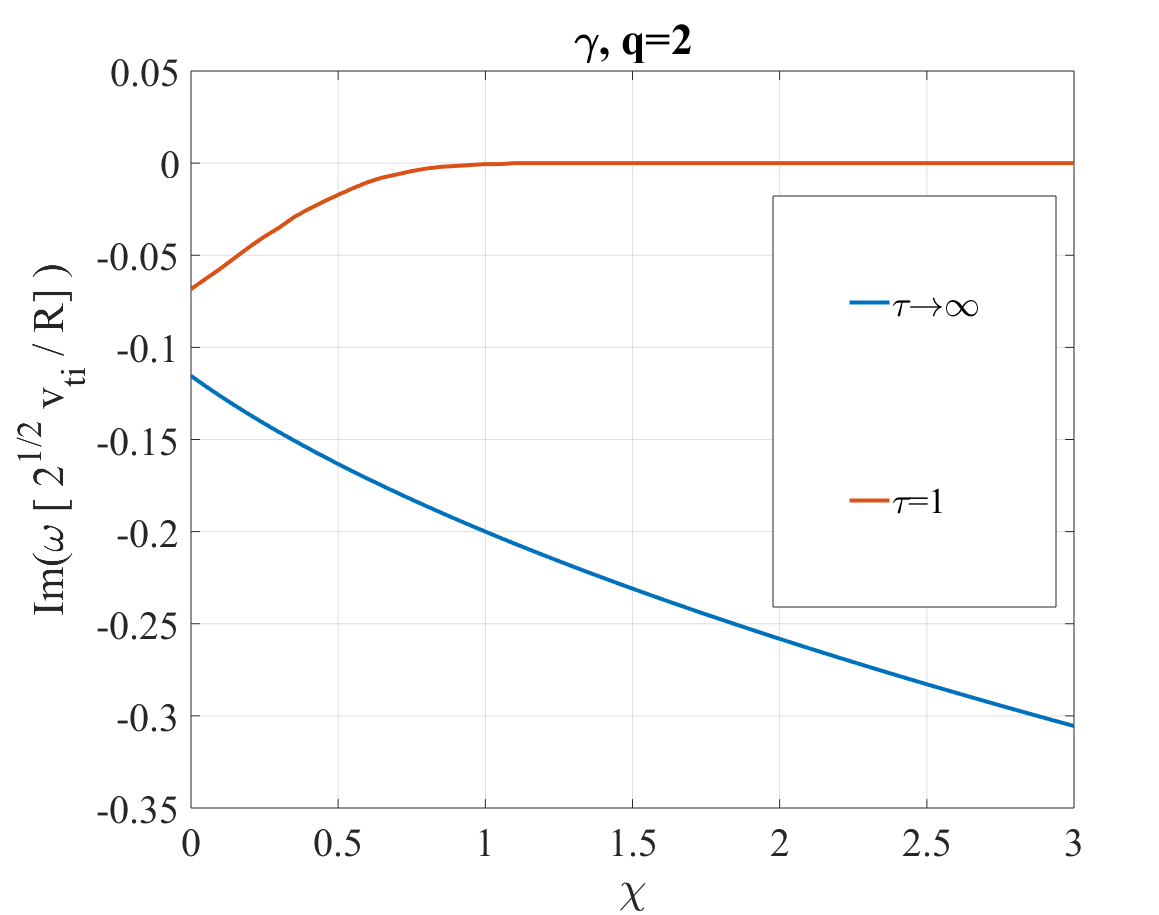}}}
    \caption{ (a)Frequency (b)Growth rate}
\end{figure}
In the following sections, we solve the complete dispersion relation (\textcolor{blue}{\ref{real}}) with realistic  values of $\tau$. 
\subsection{Effect of ion temparature anisotropy on GAM frequency and growth rate}
 \begin{figure}
    \centering
    \subfloat[\centering]{{\includegraphics[width=7.1cm]{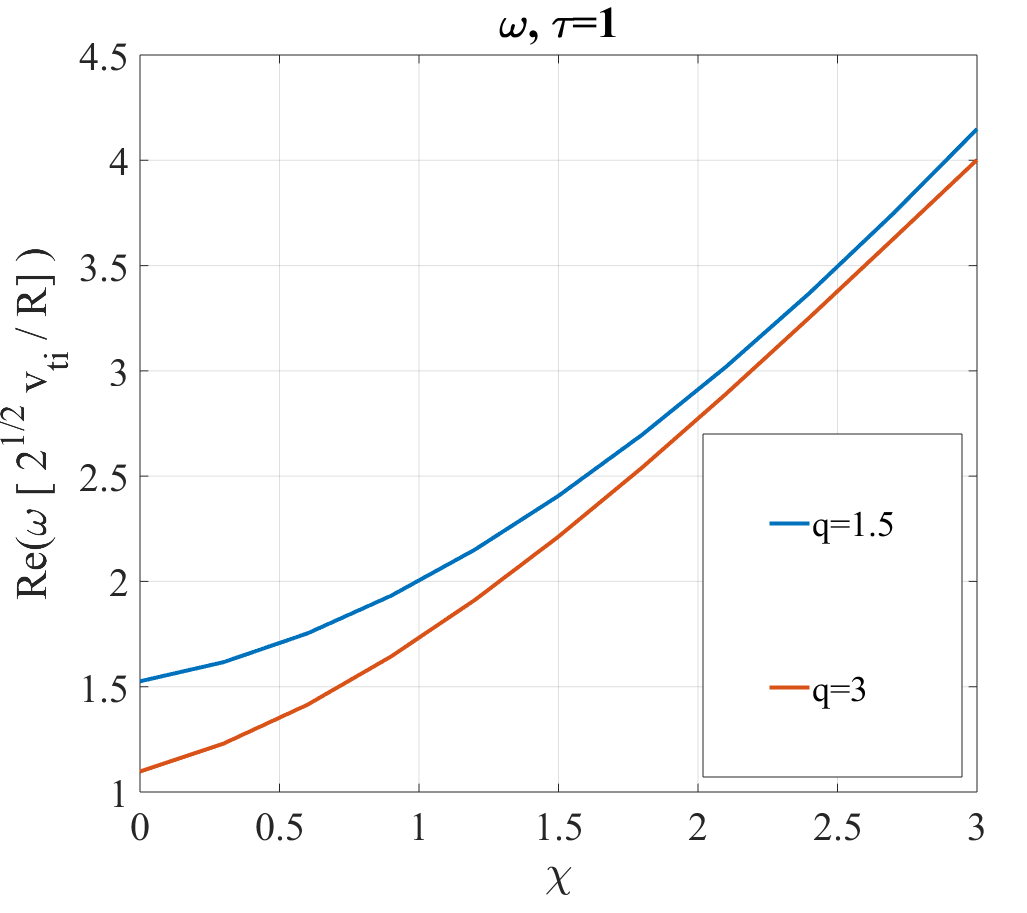}}}
    \subfloat[\centering]{{\includegraphics[width=9.5cm]{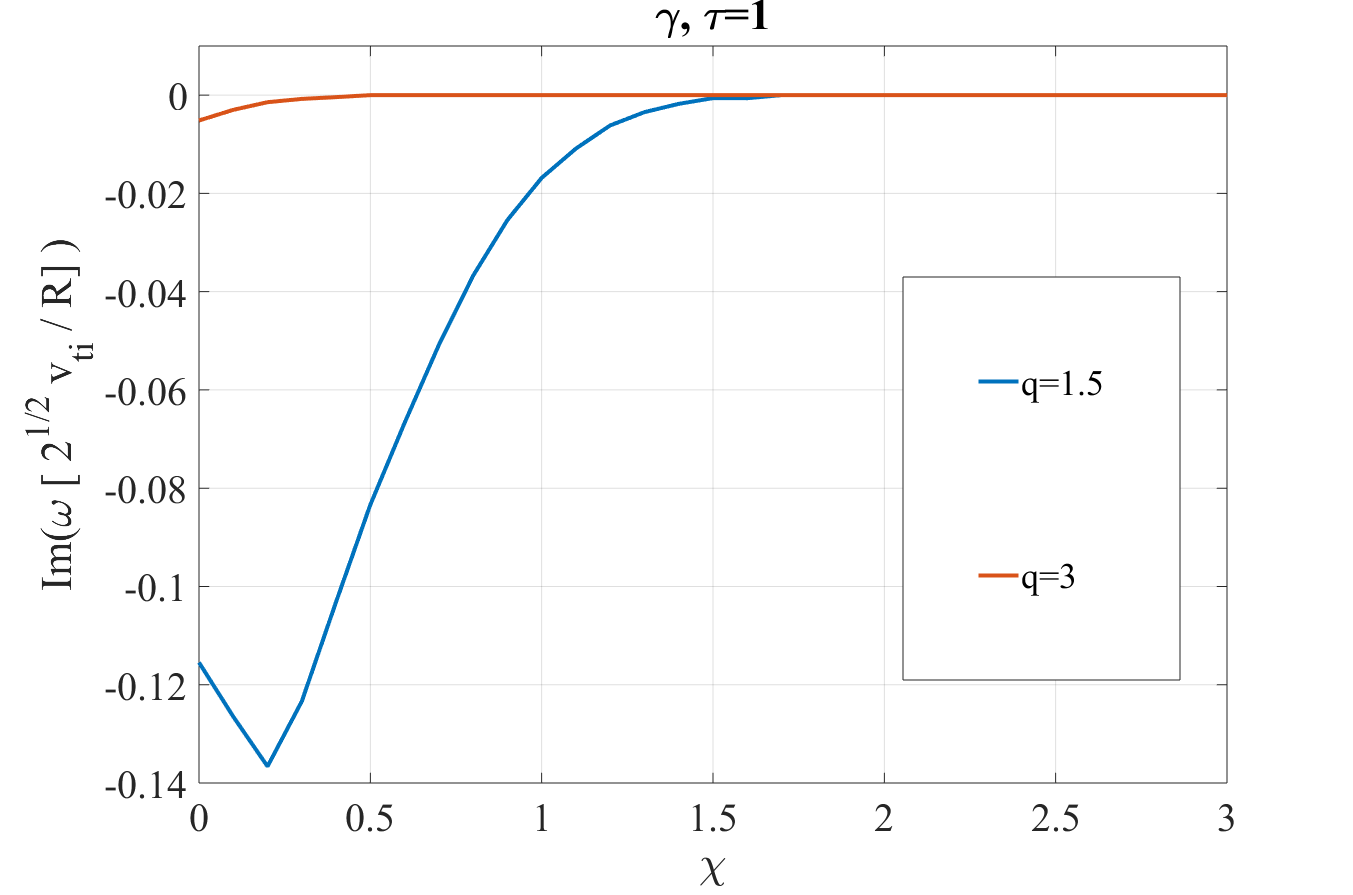}}}
    \caption{(a)Frequency (b)Growth rate}
    \label{fig}
\end{figure}
The GAM frequency and growth rate increases with $\chi$ figure \textcolor{blue}{\ref{fig}}. However, the growth rate saturate at instability margin . This saturation occurs at lower $\chi$ for higher values of $q$. It should be noted that the growth rate at higher $q$ is over estimated since damping effects due to finite orbit width is not considered in our model. These effects tend to be more important when $q$ increases \textcolor{blue}{\cite{Bc}}. Similar results were obtained in ref. \textcolor{blue}{\cite{HR}}.
\subsection{Effects of electron to ion temperature ratio}
\begin{figure}
    \centering
    \subfloat[\centering]{{\includegraphics[width=7.7cm]{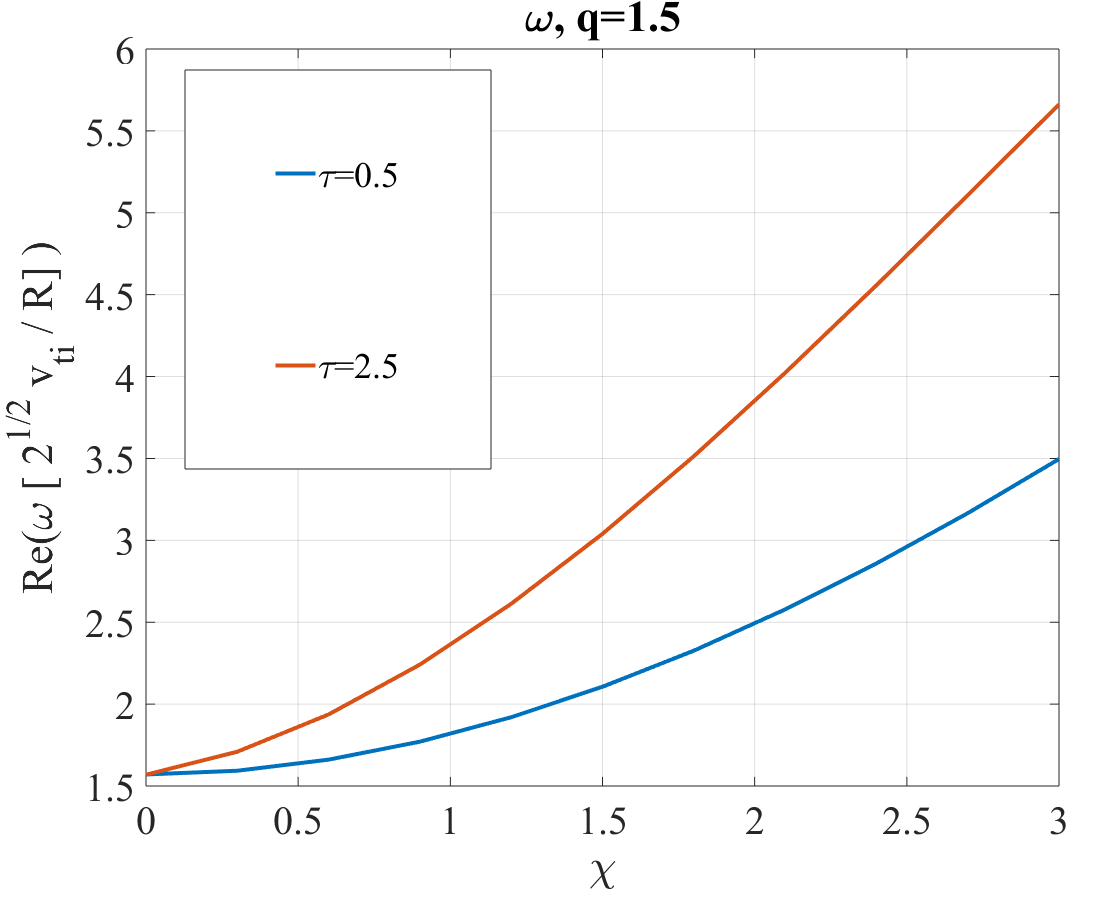}}}
    \subfloat[\centering]{{\includegraphics[width=9.5cm]{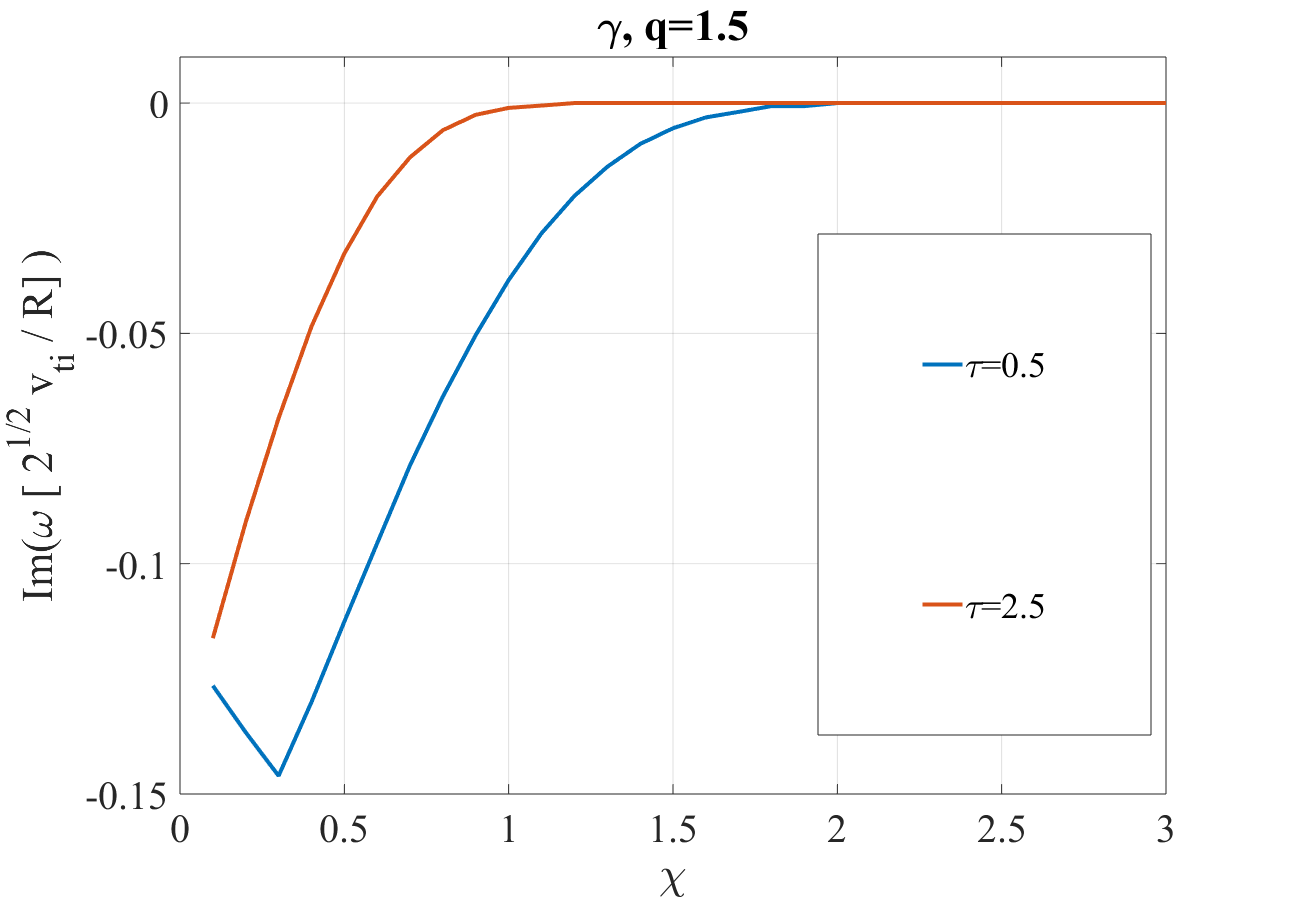}}}
    \newline
    \subfloat[\centering]{{\includegraphics[width=8.0cm]{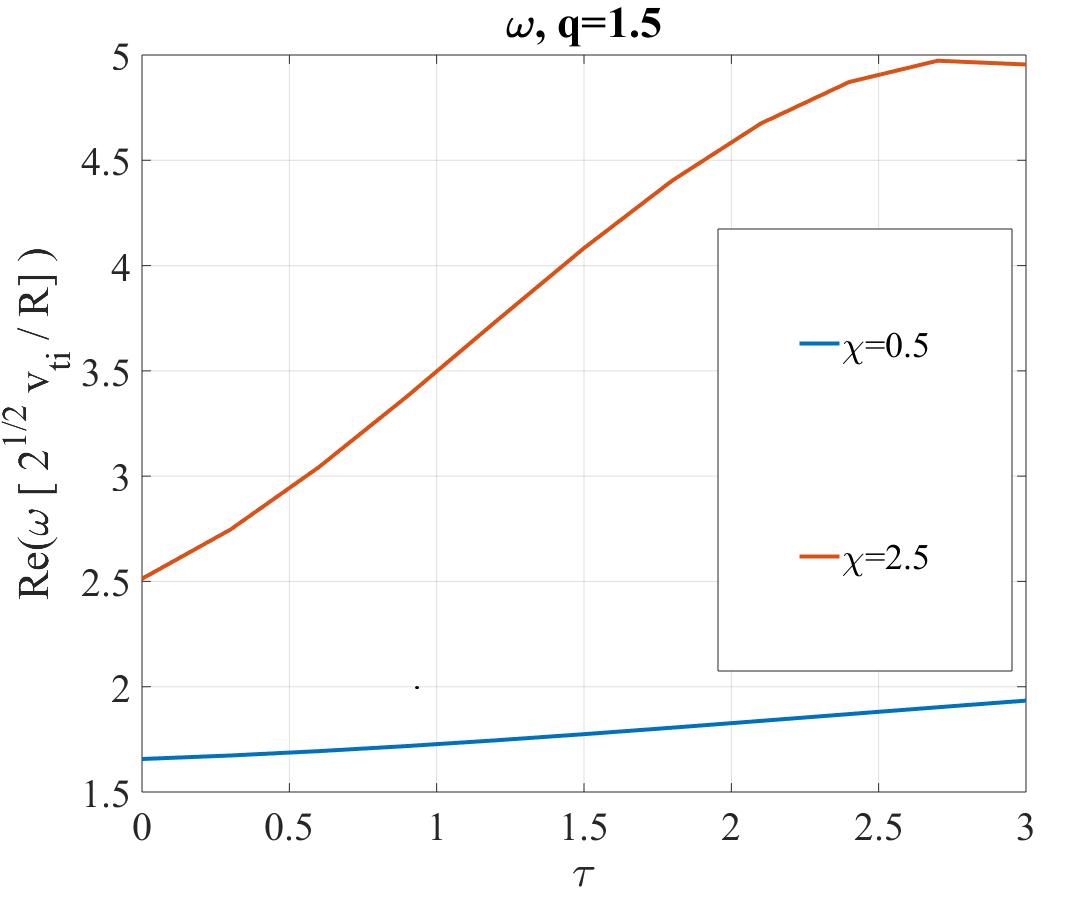}}}
    \subfloat[\centering]{{\includegraphics[width=9.1cm]{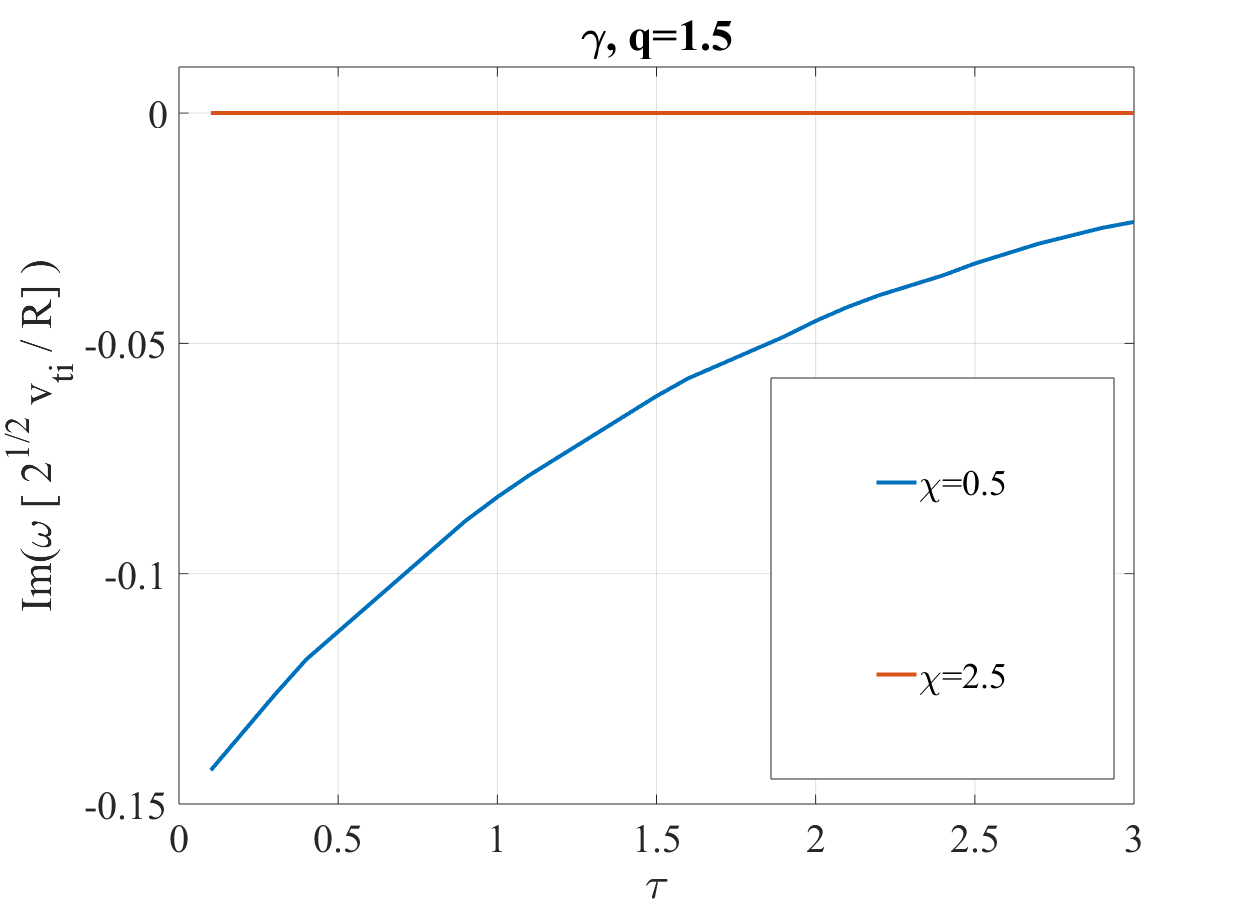}}}
     \caption{Effects of $\tau$}
      \label{fh}
    \end{figure}
    \begin{figure}
    \subfloat[\centering]{{\includegraphics[width=8.0cm]{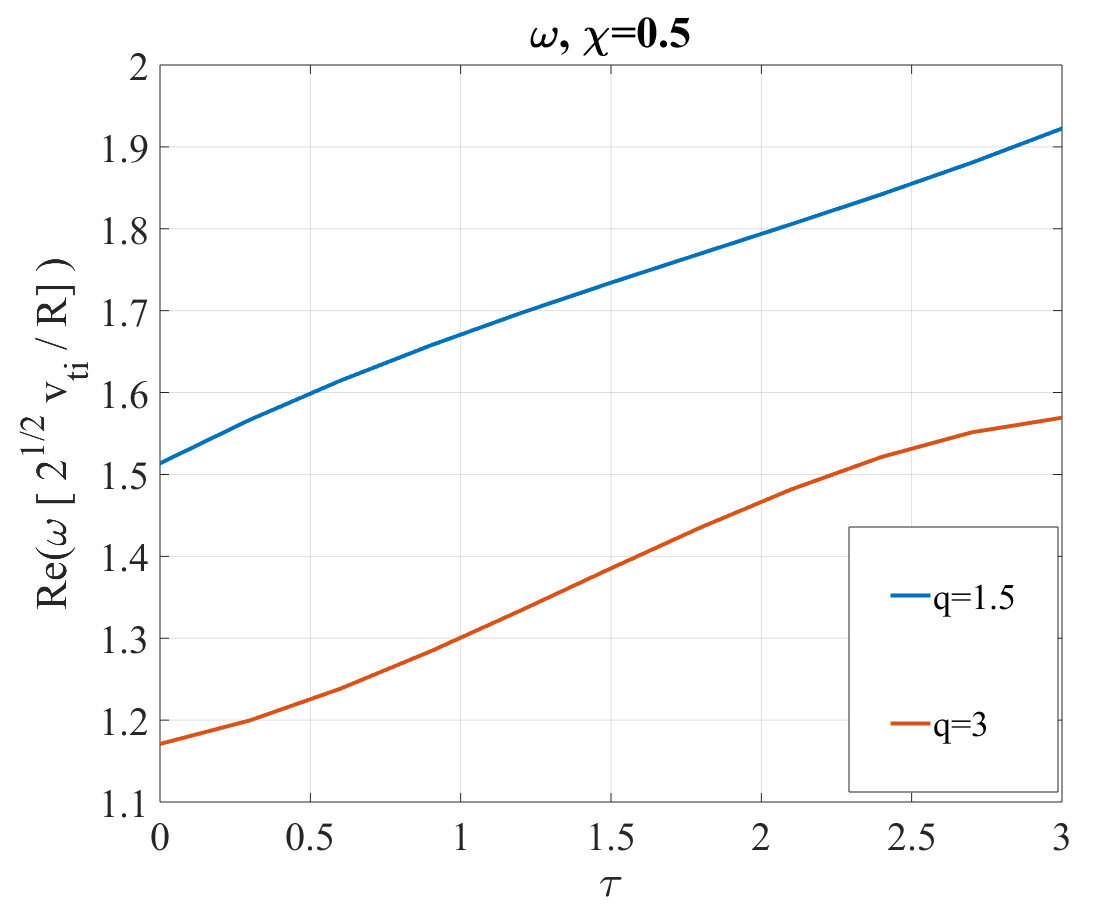}}}
     \subfloat[\centering]{{\includegraphics[width=9.5cm]{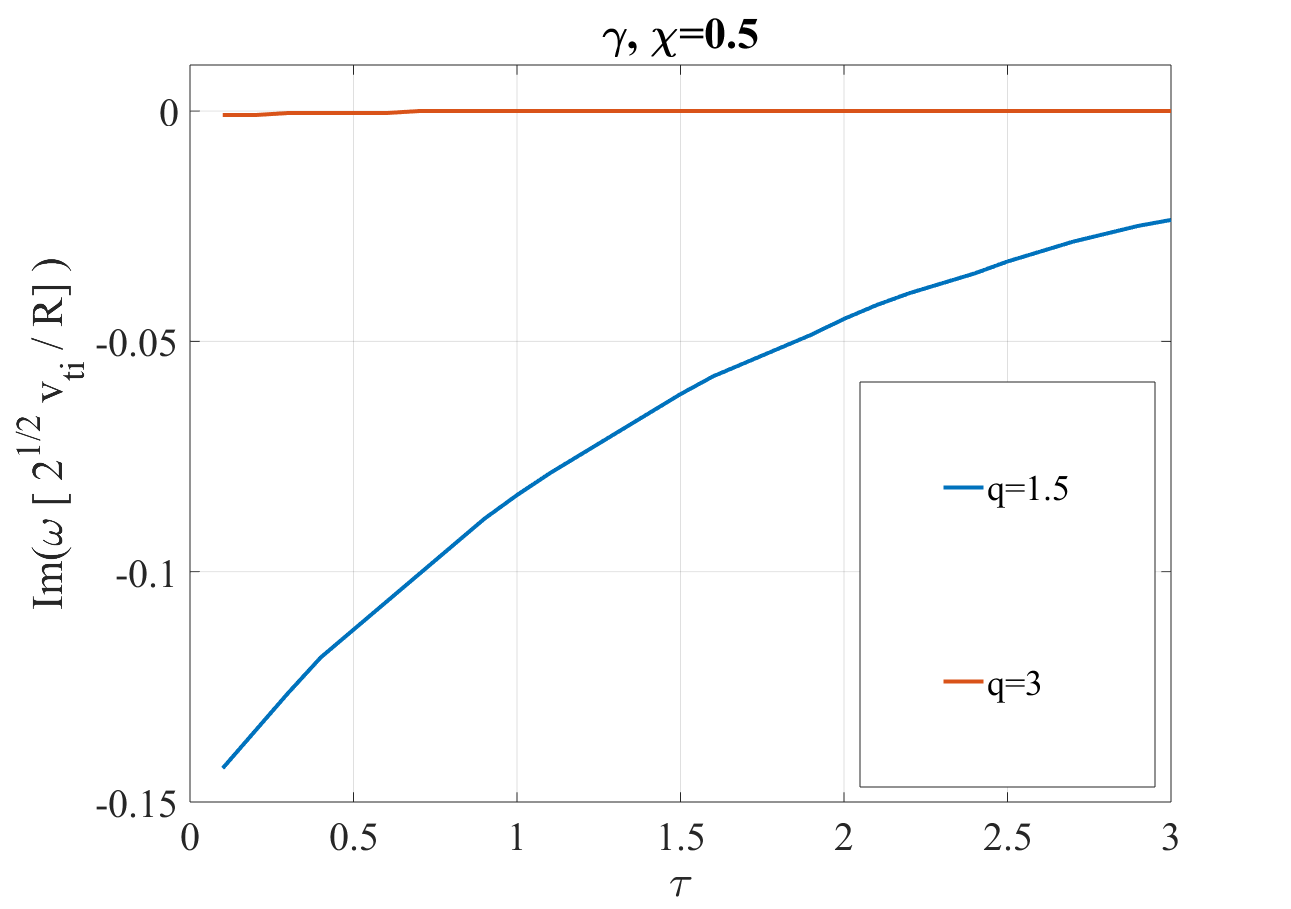}}}
    \caption{Effects of $\tau$}
    \label{fig1}
\end{figure}
In this section, we study the effect of a finite $\tau$ (this parameter was neglected in ref. \textcolor{blue}{\cite{HR}}). A summary of the  results obtained are shown in figures  \textcolor{blue}{\ref{fh}}, \textcolor{blue}{\ref{fig1}}. GAM frequency and growth rate significantly increases with $\tau$ as shown on these figures. Neglecting this parameter can lead to an underestimation of the frequency and an overestimation of the damping rate. 
\section{Application to an experimentally relevant case}
In this section, we apply the theory we have developed to an experimentally feasible case. For simplicity we consider the case where the electron to ion temperature ratio is one ($\tau=1$). K. Sasaki in  ref.\textcolor{blue}{\cite{K}} studied ion temperature anisotropy in the reverse field pinch device EXTRAP-T2. In that article $\chi\sim 0.5$ was measured. We assume such values of $\chi$ are feasible in tokamaks. To plot the the GAM frequency spectrum, we use the safety factor profile from NLED-AUG \textcolor{blue}{\cite{GV}}.
\begin{figure}[H]
    \centering
    \subfloat[\centering]{{\includegraphics[width=9.0cm]{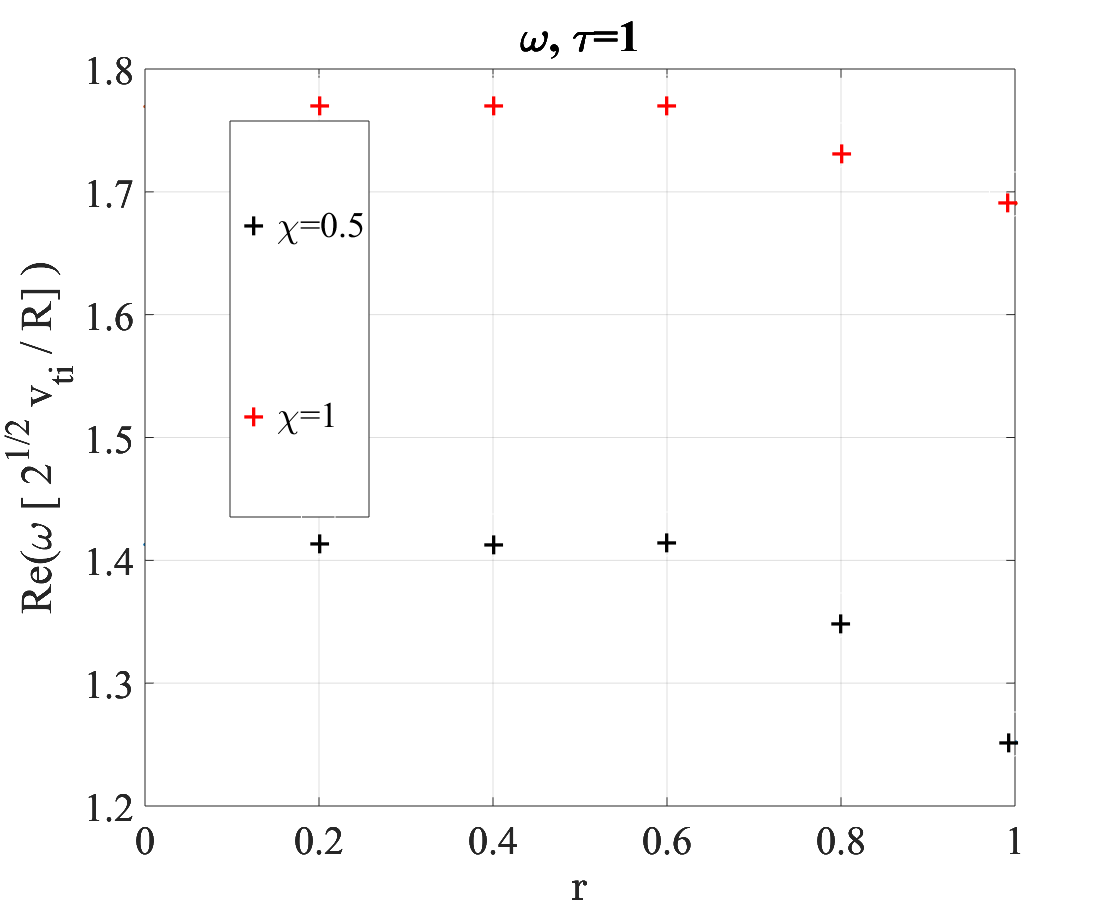}}}
    \subfloat[\centering]{{\includegraphics[width=9.0cm]{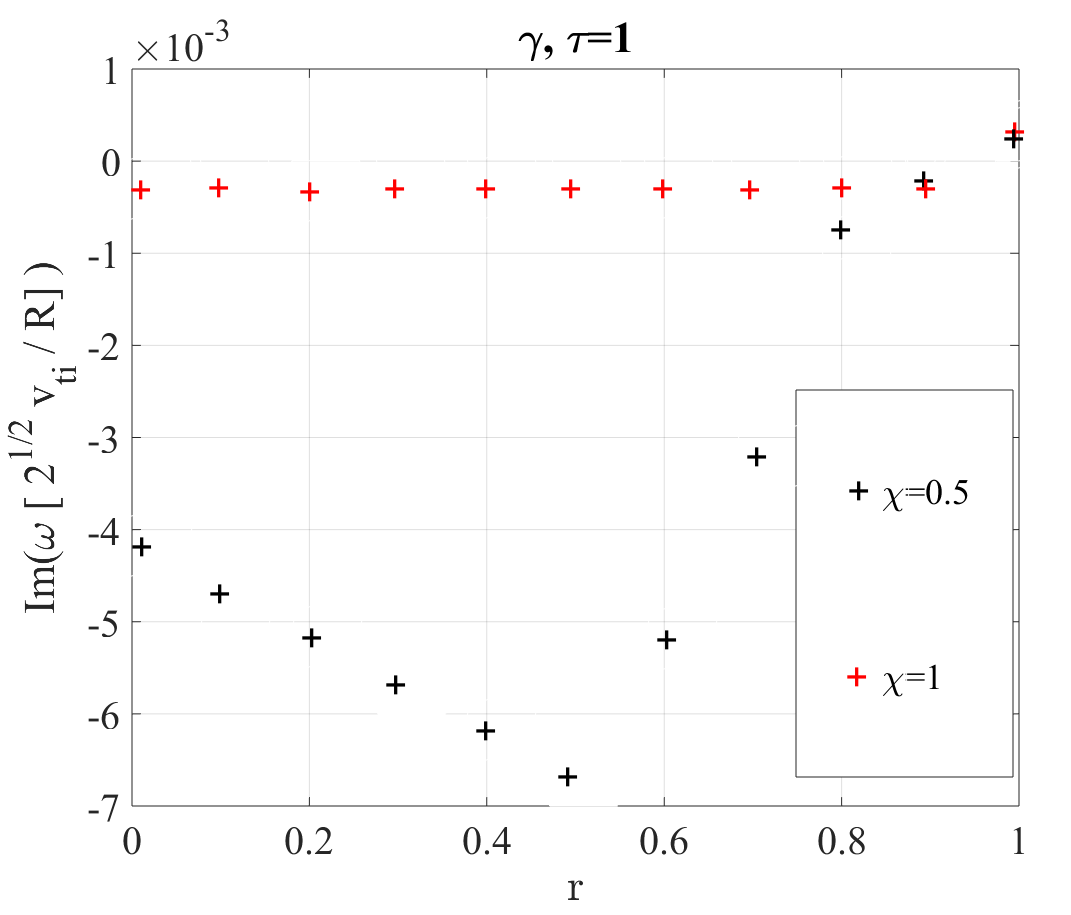}}}
    \newline
    \caption{(a)Frequency versus radial position ($r$), (b)Growth rate versus radial position ($r$)}
    \label{fig5}
\end{figure}
\begin{figure}[H]
 \centering
\includegraphics[width=0.5\textwidth]{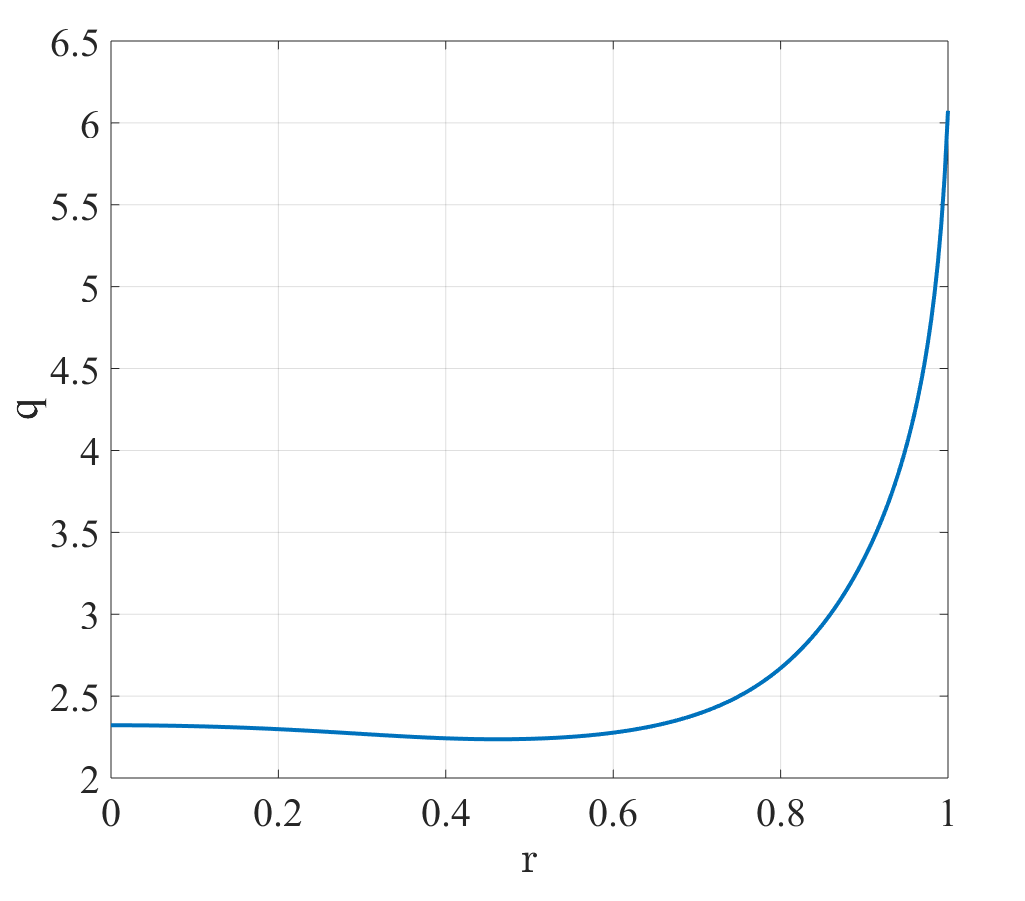}
\caption{Safety factor profile}
\label{fig6}
\end{figure}
We observe from these graphs that the frequency spectrum of GAM is very sensible to the safety factor in the presence of ion anisotropy. This is particular true closer to the plasma core where the GAM damping rate is almost an order of magnitude higher than in the isotropic case (damping due to finite orbit width which is important at higher values of $q$ is not considered in this work). Even though the GAM are heavily damped in the core in the presence of anisotropy as shown on the graph, the core dynamic of GAMs is however important as  it can significantly modify the interaction of GAMs and energetic particle in the core, the so called EGAMs \textcolor{blue}{\cite{fv,br}}.
\section{Conclusion}
Zonal structures are axisymetric perturbations that are non-linearly generated by turbulence in fusion plasmas. There are two types i.e the zero frequency zonal flow (ZFZF) and its finite frequency counterpart the geodesic acoustic mode (GAM). GAMs are unique to configurations with closed magnetic field lines with a geodesic curvature like tokamaks.
GAM frequency is of the order of the ion sound frequency and it major damping mechanism is collisionless damping (ion Landau damping). They are of interest to future magnetic fusion devices due to their potential abilities to regulate microscopic turbulence and its associated heat and particle transport.

In this work, we revisited the linear gyro-kinetic theory of GAMs with adiabatic electrons described by a Maxwellian distribution function and kinetic ions with a gyro-tropic temperature modeled with a bi-Maxwellian distribution. We extended the linear GAM theory with an anisotropic ion temperature to include a general value of electron to ion temperature ratio and derived a general linear dispersion relation for an arbitrary ion distribution function. We showed that, in the appropriate limit, we recover the published GAM dispersion relation in ref. \textcolor{blue}{\cite{HR}}. Solving the dispersion relation for the GAM frequency and damping rate, we found that ion temperature anisotropy  yields sensible changes to GAM frequency and damping rate as both tend to be increasing functions of $\chi=\frac{T_{\perp,i}}{T_{||,i}}$. The ion Landau damping is confirmed to be stronger for smaller values of the safety factor. These observations become more pronounced when a finite electron to ion temperature is considered. The frequency and growth rate for a given $\chi$, increases significantly with $\tau=\frac{T_e}{T_i}$. Hence the effect of $\tau$ on GAM dynamics is not negligible and must be included in a complete model.

We applied our theory to an experimentally feasible scenario using the safety factor profile from NLED-AUG and assuming ion temperature anisotropy in tokamaks is close to those measured in the reversed field pinch device EXTRAP-T2. Plotting the frequency spectrum and the damping rate as function of position, we find that in this scenario, the core dynamics of GAMs is significantly modified in the presence of ion temperature anisotropy. Such a modification of the core dynamics of GAMs can impact the the interaction of GAMs and energetic particle in the plasma core.

In future works, we shall extend this linear theory of GAMs with ion anisotropy in order to study the interaction of GAMs and energetic particle in the plasma core (EGAMs), since we have seen that the core dynamics of GAMs in relevant experimental scenario is significantly modified by the ion temperature anisotropy.
\section*{Acknowledgement}
This work was partially supported by the “Lorraine Université d’Excellence”  Doctorate  fundings  (project R01PKJUX-PHD21)  belonging to the Initiative "I-SITE LUE" and R\&D Program through Korea Institute of Fusion Energy (KFE) funded by the Ministry of Science, 
ICT and Future Planning of the Republic of Korea (No. KFE-EN2241-8). Part of this work has been carried out within the framework of the EUROfusion Consortium, funded by the European Union via the Euratom Research and Training Programme (Grant Agreement No.$101052200$ EUROfusion). Views and opinions expressed are however those of the authors only and do not necessarily reflect those of the European Union or the European Commission. Neither the European Union nor the European Commission can be held responsible for them. We also thank the "Maison de la simulation Lorraine" for partial time allocation on the cluster Explore (Project No. 2019M4XXX0978). Simulations on this work were also performed on the MARCONI FUSION HPC system at CINECA. The authors are grateful to Etienne Gravier and Maxime Lesur (IJL, Nancy, France), Guilhem Dif-Pradalier and Xavier Garbet (IRFM, CEA, France), and William Bin and Stefan Schmuck (ISTP, CNR, Italy), Fulvio Zonca (CNPS, Frascati, Italy), Xin  Wang and Zhixin Lu (IPP, Garching, Germany) for useful discussions and remarks.
\bibliographystyle{ieeetr}
\bibliography{biblio.bib}
\nocite{*}


\end{document}